# Spontaneous Symmetry Breaking in the Phase Space

Y. Contoyiannis[1,5], S.G. Stavrinides[2], M. Kampitakis[3], M.P. Hanias[4], S.M. Potirakis[1] and P. Papadopoulos[1]

[1]Department of Electrical and Electronics Engineering, University of West Attica, Athens, Greece.
[2]School of Science and Technology, International Hellenic University, Thessaloniki, Greece.
[3]Major Network Installations Department, Hellenic Electricity Distribution Network Operator SA, Athens, Greece,
[4]Physics Department, International Hellenic University, Kavala, Greece.
[5]Physics Department, University of Athens, Athens, Greece.

E-mails: yiaconto@uniwa.gr (Y.C.); s.stavrinides@ihu.edu.gr (S.G.S.); m.kampitakis@deddie.gr (M.K.); mhanias@physics.ihu.gr (M.P.H.); spoti@uniwa.gr (S.M.P.); ppapadop@uniwa.gr (P.P)

**Abstract**

In this brief, the spontaneous symmetry breaking (SSB) of the $\varphi^4$ theory in phase space, is studied. This phase space results from the appropriate system of Poincare maps, produced in both the Minkowski and the Euclidean time. The importance of discretization in the creation of phase space, is highlighted. A series of interesting, novel, unknown behaviors are reported for the first time; among them the most characteristic is the change in stability. In specific, the stable fixed points of the $\varphi^4$ potential appear as unstable ones, in phase space. Additionally, in the Euclidean-time phase space a unique instability in the position of the critical point, can be created. This instability is further proposed to host tachyonic field in Euclidean space.

## 1. Introduction

There are many physical systems, the dynamics of which can be reduced to one- or two-dimensional Poincaré maps [1]. Hence, a system possessing many degrees of freedom can be described using the low dimensional Poincaré maps. According to techniques already presented in the past [2,3], this can be achieved by beginning from the differential (or integrodifferential) equations for the system-constituents or its macroscopic variables, for which a reduction process has been developed.

It is apparent that the most important advantages of creating such maps are (a) numerical processing, and (b) the emergence of unknown properties of the systems in phase space. The proper topological space that comprises such information is phase space. Phase space is the proper domain for studying saddle point, as in the cases of phase transitions and critical phenomena. Signatures of topological phase transitions are presented in [4] where topological symmetry breaking phenomena for structural solids based on the quantum spin hall effect with $C_6$ hexagonal symmetries appear.

In the present paper, the successful creation of two dimensional (2D) Poincaré maps, utilizing the Ginzburg-Landau (G-L) free energy is presented. Subsequently, it is examined whether this system of maps is consistent with the Theory of Critical Phenomena, from the phase space point of view. Similar works within the G-L theory have been presented in the past and were based on the G-L Hamiltonian, further focusing on the similarities between critical phenomena and chaotic dynamics [1].

In figure 1, the G-L free energy, when Spontaneous Symmetry Breaking (SSB) happens, as well as the inverse G-L free energy (IGLFE), are illustrated, both in the framework of the $\varphi^4$ theory. In the case of describing the symmetric phase of a second order phase transition, the G-L free energy is given by: $U(\varphi) = \frac{1}{2}r_0\varphi^2 + \frac{1}{4}u_0\varphi^4$, $(r_0 > 0, u_0 > 0)$; in this phase when $\varphi = 0$ and $U(\varphi) = 0$ the system exists in the ground state. According to the $\varphi^4$ theory, symmetric breaking is accomplished when $r_0$ changes sign, becoming negative, while $u_0$ remains positive. This SSB is presented in figure 1 with the blue curve, where the two stable fixed points are the new ground states at positions $\varphi^* = \pm\sqrt{\frac{r_0}{u_0}}$, ($r_0 = u_0 = 1$). The demonstrated free energy (blue curve), is then provided by:

$$U_{SSB}(\varphi) = -\frac{1}{2}|r_0|\varphi^2 + \frac{1}{4}u_0\varphi^4 \ (u_0 > 0). \tag{1}$$

Actually, in nature the system can exist in only one of these ground states. Thus, the symmetry appears to be broken. This phenomenon is what is called the "Spontaneous symmetry Breaking" (SSB) [5]. Note that an extension of the SSB within the Goldston Theory [6], in the three dimensions, is what is called the "Mexican hat GL free energy".

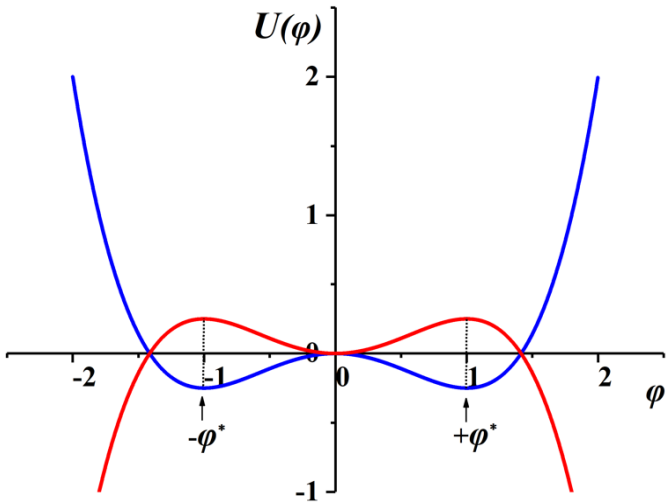

**Figure 1**. The SSB and the IGLFE are presented with blue and red color, respectively, as described in the text. The maxima and minima in the red and blue curve, respectively, are marked at $\varphi^* = \pm\sqrt{\frac{r_0}{u_0}}$, with $r_0 = u_0 = 1$.

The IGLFE ($r_0 > 0$, $u_0 < 0$) is provided by:

$$U_{in}(\varphi) = \frac{1}{2}r_0\varphi^2 - \frac{1}{4}|u_0|\varphi^4 \ (r_0 > 0), \tag{2}$$

and it is presented with the red curve in figure 1; now, two unstable points appear, in the position of the stable points. The IGLFE theory, which is expressed through equation (2), can be considered as an intermediate theory, since self-interactions appear up to $\varphi^4$, but the alteration of the sign in the $\varphi^4$ term is implemented the same way as in the $\varphi^6$ theory.

The scope of this work is to transfer relations (1) and (2), coming from critical theories, in topological spaces as the phase space of Poincare maps, in order to research for an another description of the SSB phenomenon. As it is known [7], there are soliton-solutions in the SSB theory, which live in Minkowski's space, and instanton-solutions in the IGLFE, which live in the Euclidean space. So, in this work we will attempt to produce Poincare maps and their phase space in these two spaces (Minkowski and Euclidean).

## 2. Construction of Poincare maps

The critical configurations are produced through the saddle-point approximation and the Euler-Lagrange (E-L) equation of the motion [8,9], in a parametric $\lambda$-space, is provided by the following equation:

$$\frac{\partial^2\varphi}{\partial\lambda^2} = -\frac{\partial}{\partial\varphi}U(\varphi). \tag{3}$$

Utilizing equation (3) the 2D Poincaré map is produced. The reduction to a discrete map takes place within the parametric space of $\lambda$. This reduction is achieved via the transition $n \to n + 1$, where $n$ stands for the length of the map trajectory in time. As a result, parameter $\lambda$ should be connected to the map trajectory of length $n$. A way to achieve this is by utilizing the inductive relations, known as generator-functions, of the renormalization group Hu-Rudnick [10]. The resulting linear relation between $\lambda$ and $n$ is considered in the following equations and equation (5) is a Hu-Rudnick type relation, $k$ being a scaling factor:

$$\lambda_n = kn \tag{4}$$

$$\lambda_{n+1} = kn + k \Longrightarrow \lambda_{n+1} = \lambda_n + k. \tag{5}$$

This way, discretization can be achieved by replacing in equation (3) parameter $\lambda$ as in the following relation:

$$\lambda = kn, \tag{6}$$

rendering parameter $\lambda$ proportional to time.

By replacing in equation (3) $U(\varphi) = U_{SSB}(\varphi)$, since this produces soliton-solution in Minkowski's space, then:

$$\lambda_M = kn_M, \tag{7}$$

whereas for $U(\varphi) = U_{in}(\varphi)$, in the Euclidean space we get:

$$\lambda_E = kn_E, \tag{8}$$

where $n_M$ and $n_E$ are the discrete forms of Minkowski's and Euclidean times, respectively. Consequently, the E-L equation (3) is written, in each case respectively, as:

$$\frac{d^2\varphi}{d\lambda_M^2} = -\frac{\partial U_{SSB}(\varphi)}{\partial\varphi} \tag{9}$$

$$\frac{d^2\varphi}{d\lambda_E^2} = -\frac{\partial U_{in}(\varphi)}{\partial\varphi}. \tag{10}$$

In the case of Minkwoski's time, using the equation (7), equation (9) can be rewritten as follows:

$$\frac{d^2\varphi}{dn_M^2} = -k^2(-|r_0|\varphi + u_0\varphi^3). \tag{11}$$

The equivalent system of first order differential equations is:

$$\psi = \dot\varphi \tag{12}$$

$$\psi = -k^2(-|\dot{r}_0|\varphi + u_0\varphi^3). \tag{13}$$

By omitting the index in $n$ in the last two equations, the 2D Poincaré map can be expressed as:

$$\varphi_{n+1} = \varphi_n + \psi_n \tag{14}$$

$$\psi_{n+1} = \psi_n - k^2\varphi_n(-|r_0| + u_0\varphi_n^2), (u_0 > 0), \tag{15}$$

where $n$ is the discrete Minkowski's real time.

In an equivalent way, in the Euclidean time case and by utilizing equation (8), equation (10) can be rewritten as follows:

$$\frac{d^2\varphi}{dn^2_E} = -k^2(r_0\varphi - |u_0|\varphi^3), \tag{16}$$

which equivalently is written as the following set of equations:

$$\psi = \dot{\varphi} \tag{17}$$

$$\dot{\psi} = -k^2(r_0\varphi - |u_0|\varphi^3). \tag{18}$$

By omitting again the index in $n$ in the last two equations, the 2D Poincaré map, in the Euclidean case, can be written as:

$$\varphi_{n+1} = \varphi_n + \psi_n \tag{19}$$

$$\psi_{n+1} = \psi_n - k^2\varphi_n(r_0 - |u_0|\varphi_n^2), (r_0 > 0), \tag{20}$$

where now $n$ stands for the discrete Euclidean imaginary time.

## 3. Phase spaces and invariant densities

The notion of phase space originates form mechanics, and it is defined as the two-dimensional portrait of position and velocity (time derivative of position). A generalization of this space in maps provides with 2D maps of the form $(\varphi_n, \psi_n = \varphi_{n+1} - \varphi_n)$. Regarding the invariant density, this is an invariant measure of maps, defining the density of mapping repetitions or equivalently the distribution of $\varphi_n$.

### 3.1. In Minkowski's time

The phase diagram and the invariant measure of the maps described in equations (14) and (15) in Minkowski's time, are presented in figure 2a, for initial conditions very close to the critical point $(\varphi = 0, \psi = 0)$. In this case the parameter values are: $k^2 = 10^{-6}$, $|r_0| = 1$, $u_0 = 1$, $n_{max} = 2 \cdot 10^5$, while the initial values are set to $\varphi_1 = \psi_1 = 10^{-4}$ (very close to zero). In figure 2a the phase space diagram $(\varphi_n, \psi_n)$ is shown, while in figure 2b the corresponding graph of the invariant measure is presented.

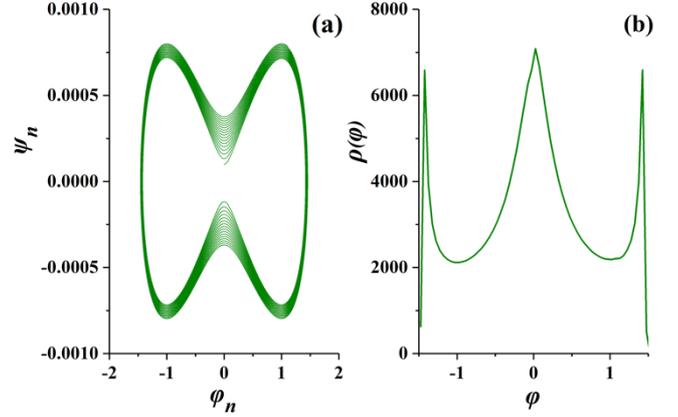

**Figure 2.** (a) The phase space diagram in Minkowski's time. (b) The corresponding invariant measure. The critical point at $\varphi = 0$ is present although we are not in a symmetric phase. Two new points, one on each side of the critical point, appear. The steep peaks indicate that in this case these points are unstable.

The following remarks, regarding figure 2 should be pointed out:

Three unstable fixed points appear (at the positions of the three peaks of figure 2b showing the invariant measure), which, as proved in the appendix, are unstable.

The peak at $\varphi = 0$ is the critical point. This peak should not appear because the symmetry has been broken.

The new unstable fixed points do not appear at the proper position at which they should appear according to theory, i.e., at the same positions at which the stable fixed points of $U_{SSB}(\varphi)$ appear, $\varphi^* = \pm\sqrt{\frac{r_0}{u_0}}$ (see also section 1 and figure 1). On the contrary, they are shifted to higher absolute values. The new symmetric positions of fixed points are the limits of a non-harmonic oscillator, which moves between the wells of the $U_{SSB}(\varphi)$. These positions belong on the separatrix line and can be defined as the intersection points between $U_{SSB}(\varphi)$ and $U_{in}(\varphi)$. As a result, the new fixed points are $(\tilde{\varphi} = \pm\sqrt{\frac{2|r_0|}{u_0}}, \tilde{\psi} = 0)$.

The two minima apearing in the invariant measure (figure 2b) indicate that at the positions of the stable fixed points of $U_{SSB}(\varphi)$ $(\varphi^* = \pm\sqrt{\frac{r_0}{u_0}}, \psi^* = 0)$, there is minimum probability of trajectory appearance (which is also obvious from figure 2a).

Note also that in the case that the initial conditions are moved away from the critical point $(\varphi = 0, \psi = 0)$, the critical fixed point in the middle is vanished and the two new symmetric unstable fixed points at $(\tilde{\varphi} = \pm\sqrt{\frac{2|r_0|}{u_0}}, \tilde{\psi} = 0)$ remain.

Since the purpose of this work was the generation of Poincare maps and the corresponding phase spaces in time, it is noted that the spatial term $\nabla^2\varphi$ in the E-L equation of motion, has been ignored. This means that the above mentioned saddle points dominate the infrared limit, i.e., longer wavelengths. Thus, keeping only equation (3), which is a common differential equation of motion, where the field $\varphi$ is a function only of time, time $t$ is expressed by parameter $\lambda$. The resulting differential equation is that of a quadratic-quartic oscillator [11] within the configuration space. At this limit, the time scale for significant variations is large and therefore the time derivative is interpreted in a coarse graining sense. Taking into account this framework, discretization in time $\lambda$ through relation (6), is feasible. Wishing to keep in touch with the continuous model, factor $k$ needs to get small values ($k \ll 1$). It is proved in the appendix that this selection is of significant importance in creating unstable fixed points, while provides with the possibility for initial conditions very close to the origin (0,0) and thus to see something that in real space is impossible to see, the appearance of an unstable critical point.

### 3.2. In Euclidean time

In figure 3 the phase diagram and the invariant measure of the maps described in equations (19) and (20), in the case of Euclidean, are presented. The parameter values are: $k^2 = 10^{-5}$, $|r_0| = 1$, $u_0 = 1$, $n_{max} = 100000$, while the initial values were set very close to zero, i.e., $\varphi_1 = \psi_1 = 10^{-5}$.

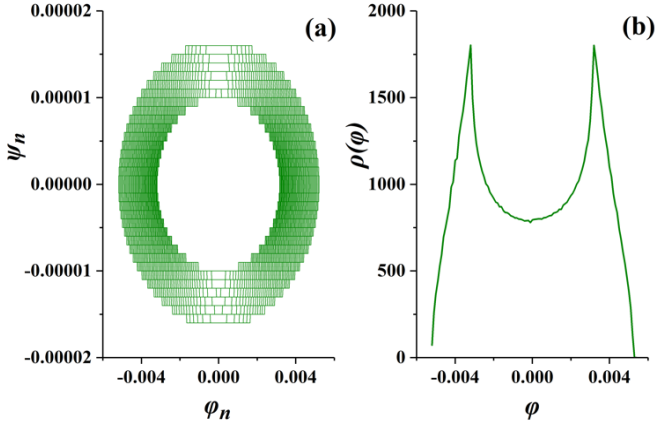

**Figure 3**. The phase space and the invariant measure for Euclidean time, $k^2 = 10^{-5}$, $|r_0| = 1$, $u_0 = 1$, $n_{max} = 100000$, while the initial values were set to $\varphi_1 = \psi_1 = 10^{-5}$ (very close to zero).

From the invariant measure results (figure 3b), the following should be pointed out:

Although the presented case is close to the critical point, in contrast to the case of Minkowski's time, in the case of Euclidean time, the third peak does not appear. In the appendix it is proved that these fixed points are unstable.

The local minima no longer exist.

By changing the parameter values in Euclidean time, an interesting phenomenon appears. When the ratio $\Lambda \equiv r_0/u_0$ increases, the two unstable fixed points move one towards the other. For a unique value of the ratio ($\Lambda_{cr}$) the two fixed points are merged to one point, located at the position (0,0), namely at the position of the critical point. As a result, for $\Lambda \geq \Lambda_{cr}$ there is only one unstable fixed point, as shown in figure 4 where for $\Lambda_{cr} = 12.5$ the symmetric unstable fixed points are vanished and a critical unstable fixed point appears.

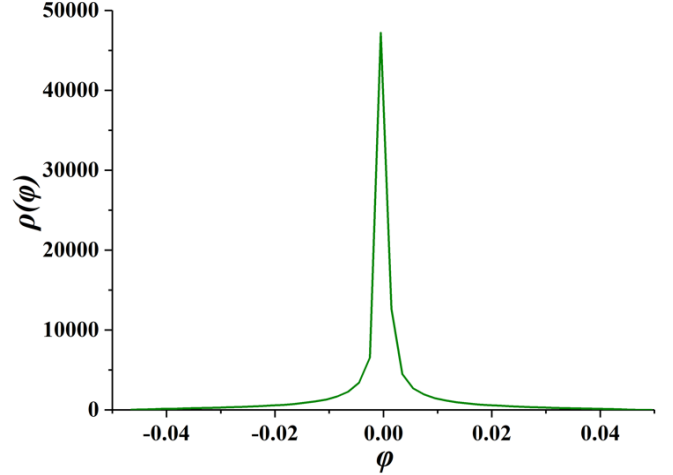

**Figure 4**. The invariant measure for Euclidean time, $k^2 = 10^{-5}$, $|r_0| = 12.5$, $u_0 = 1$, $n_{max} = 100000$, while the initial values were set to $\varphi_1 = \psi_1 = 10^{-6}$ (very close to zero).

If the initial conditions get away from the critical point (0,0) then the two unstable fixed points never merge to one point, no matter how close they are to each other. So, in this case no unstable critical fixed point exists (as in figure 4).

On the contrary, in Minkowski's space, where the positions of symmetric unstable fixed points are determined by the intersection points of $U_{SSB}(\varphi)$ and $U_{in}(\varphi)$, there is no characteristic point where the unstable points merge to. Note that the position of the unstable fixed point in figure 3 is varied, as the ratio $\Lambda$ is varied. This dependence is further described by the power-law, described by the following equation:

$$\varphi^* = f\Lambda^{-0.5}. \quad (21)$$

The plot of this power law appears in figure 5.

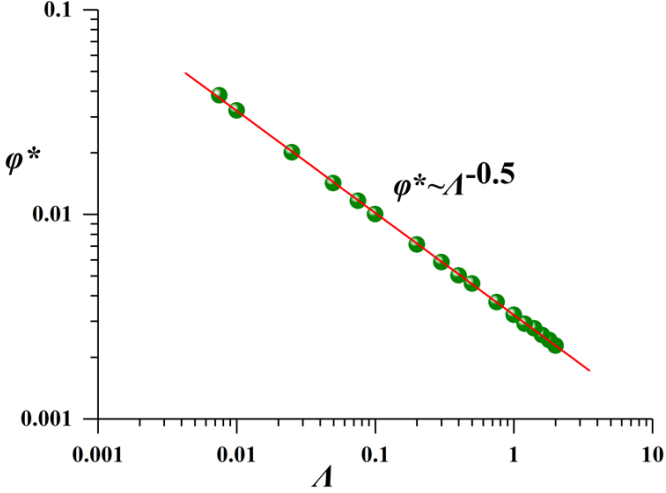

**Figure 5**. The value of unstable fixed points $\varphi^*$ in Euclidean time vs. the ratio $\Lambda$. A power-law $\varphi^* = f\Lambda^{-0.5}$ is revealed for $k^2 = 10^{-5}$, $n_{max} = 100000$, $\varphi_1 = \psi_1 = 10^{-6}$, while we have considered $f = 0.21$.

## 4. Discussion about some possible explanations on our findings

In this section we provide some possible explanations about the observed appearance of the unstable critical point during the phase of broken symmetry, as well as for the shift of the positions of the new fixed points at absolute values higher than those predicted from theory.

The map considered for the case of Minkowski's time (see equations (14) and (15) in section 2) describes the dynamics of the order parameter just after the symmetry breaking takes place. At this point, we would like to stress that in the case of Minkowski's time presented in section 3.1 (figure 2) the saddle point character of the critical point is responsible for the appearance of the unstable critical point.

Very close to the critical point, $\varphi = 0$, the trajectories in a narrow area around $\varphi = 0$ "perceive" both the unstable and the stable dynamics of the saddle point. Due to the fact that the width of this area is small in order to observe its existence, one has to choose initial conditions very close to $\varphi = 0$ and in addition to investigate the dynamics by using a fine discretization, i.e., small scaling factor $k$. As a result, in this regime, a peak at $\varphi = 0$ emerges which possesses the form of a cusp, reflecting the fact that $\varphi = 0$ is unstable in the SSB phase. If one starts the trajectories at a larger distance from $\varphi = 0$ or/and considers the dynamics for larger $k$, then the peak at $\varphi = 0$ is not visible anymore. The necessity for the existence of the unstable critical point until the completion of SSB is explained in detail in [12,13], where we have studied the SSB phenomenon for the Ising thermal model within the frame of critical phenomena theory (not in phase space). In these papers, we have shown that SSB is completed as soon as the "comunication" between the two degenerated vacua is no more possible, i.e., when the two degenerated vacua are completely separated (green curve in figure 1 of [12]). This is achieved as soon as the control parameter, i.e., temperature $T$, gets reduced by a small amount $\Delta T$ below its critical value $T_c$. Within the narrow band of temperatures $\Delta T$ the "communication" between the two degenerated vacua is possible due to the existence of the unstable critical point that repels the trajectory from one vacuum to the other. As soon as the SSB is completed and the two completely independent vacua are formed, then the unstable critical point has no role to play any more. Therefore, after the complete separation, for temperatures lower than $T_c - \Delta T$, one cannot observe the unstable critical point but only the two new independent fixed points.

On the contrary, in the case of the Euclidean time (section 3.2, figure 3) the unstable trajectories at the critical point are not present. The reason is that in this case the critical point has not saddle point behavior. As a result, contrary to the case of Minkowski's time, no unstable critical point appears at the position (0,0). Nevetheless, as we have shown in section 3.2, for a specific value of the ratio $\Lambda$ an unstable critical point does appear at the position (0,0). However, in this case its existence is not due to saddle point dynamics but due to the approach of the two symmetric unstable fixed points until they are merged into one point at (0,0) (see figure 4).

The new fixed points at the phase diagram, after the SSB, do not appear, as already mentioned, at the positions that SSB potential theory predicts (see figure 1) but appear shifted to higher absolute values. The form of the invariant measure with 3 peaks (as in figure 2) implies the existence of $\varphi^6$ potential (tricritical point, see reference [5]). The difference in this case is that this form is not a result of $\varphi^6$ potential but emerges only when one approaches very close to the critical point, where the trajectories "sense" the unstable saddle point character of the critical point, as previously mentioned. At the limit where the initial values tend to the critical point, the symmetric new fixed points tend to be at the positions $\pm\sqrt{\frac{2|r_0|}{u_0}}$, as already mentioned. These positions are found in the interval between the positions $\pm\sqrt{\frac{r_0}{u_0}}$ of the fixed points of the potential $U_{SSB}$ for the second order phase transitions and the positions of the fixed points $\pm 2\sqrt{\frac{r_0}{|u_0|}}$ of the $\varphi^6$ potential for the first order phase transitions. This observation may lead us to the conclusion that this intermediate position of the symmetric fixed points is due to the existence of an "intermediate" potential which could be an effective potential, appearing as a result of special conditions, namely, (a) for initial conditions very close to the critical point, and (b) for fine discretization (that keeps the trajectory for longer time close to the critical point). Thus, one could consider the existence of an effective potential that determines the dynamics of the trajectory. Therefore, the shift of the fixed points to higher absolute values could be a result of the existence of this effective potential, which is something worth-investigating in the future.

## 5. Discussion about the tachyonic field in phase space

In field theories parameter $r_0$ can be consider as the $m^2$ quantity, where $m$ is the mass term in the Klein Gordon wave equation, which can describe the scalar tachyons [7]. Consequently, when the symmetry is broken this term ($r_0 = m^2$) becomes negative; thus, the mass becomes an imaginary quantity. This is the tachyonic mass [7]. For a particle such as a tachyon, possessing an imaginary mass means that this particle (the tachyon) moves with a speed higher than the light speed.

Due to the fact that $m^2 < 0$ in case of broken symmetry phase, it becomes apparent that the tachyonic field appears in this phase, i.e., the broken symmetry phase. It is known that the tachyon is a very unstable state [14-19]. So, it seems that the emerging instabilities at the critical point could be proposed for hosting the tachyons. This is further supported by the fact that tachyons field in Minkowski's space live and appear when the SSB phenomenon takes place.

As already mentioned above, the SSB can described by Poincaré maps in Euclidean time. This allows for introducing a type of tachyons field that lives in Euclidean space. In Mincowski's time, tachyons violate the principle of causality. But this is not the case when we refer to Euclidean time, where such a problem does not exist. Euclidean time is the inverse temperature and this way no violation of causality exists. Thus, instabilities in the critical point, as in case illustrated in figure 4, could support the appearance of tachyons in imaginary time (or the inverse-temperature space) [20]. In this work we introduce this idea for further investigation.

## 6. Conclusions

The discretization we impose on the equation of motion, resulting from the Ginzburg-Landau potentials that describe the SSB in $\varphi^4$ theory, changes the stability of the system by converting the stable points into unstable points in the phase space. These instability effects appear in both the Minkowski's time, as well as in the Euclidean time. For initial conditions very close to the critical point, in Minkowski's time the critical point still exists even in the phase of SSB, becoming an unstable fixed point. In the case of Euclidean time, the existence of unstable critical point is due to the approach of the two symmetric unstable fixed points until they are merged into one point. The introduction of a type tachyon field living in Euclidean time is favored from such an instability. An extension of the present work in $\varphi^6$ critical theory of the first order phase transition is within our future plans.

## Acknowledgements

The authors would like to thank Prof. F. Diakonos (Physics Department, University of Athens) for his constructive comments and suggestions, during the preparation of this work.

## Appendix

### Calculation of Lyapunov exponents

*A. In Minkowski's time*

Initially, we calculate the fixed points $\varphi^*, \psi^*$ of the maps described in equations (14) and (15), which are in Minkowski's time. They can be found by solving the system:

$$\varphi^* = \varphi^* + \psi^* \tag{A1}$$

$$\psi^* = -k^2 \varphi^* (-|r_0| + u_0 \varphi^{*2}) + \psi^*. \tag{A2}$$

It is apparent that the stability of the fixed points is determined by the eigenvalues $\lambda_{L_{1,2}}$ of Liapunov exponent $\lambda_L$ of the characteristic equation:

$$|J - \lambda_L I| = 0, \tag{A3}$$

where $I$ is the identity matrix and $J$ is the Jacobian matrix, appearing below:

$$J = \begin{pmatrix} 1 & 1 \\ k^2 |r_0| - 3k^2 u_0 \varphi^{*2} & 1 \end{pmatrix}. \tag{A4}$$

Then, equation (A3) can be written as:

$$\begin{vmatrix} 1 - \lambda_L & 1 \\ k^2 |r_0| - 3k^2 u_0 \varphi^{*2} & 1 - \lambda_L \end{vmatrix} = 0 \Rightarrow$$

$$\lambda_L^2 - 2\lambda_L + (1 - k^2 |r_0| + 3k^2 u_0 \varphi^{*2}) = 0, \tag{A5}$$

resulting into the following solutions for the exponent $\lambda_L$:

$$\lambda_{L_{1,2}} = 1 \pm k(|r_0| - 3u_0 \varphi^{*2})^{1/2}. \tag{A6}$$

At the critical point ($\varphi^* = 0, \psi^* = 0$), equation (A6) provides with: $\lambda_{L_{1,2}} = 1 \pm k(|r_0|)^{1/2}$. Since the value of the scaling parameter $k$ is very small, $\lambda_L > 0$. As a result, the fixed point on the critical position is unstable. Higher values of $k$ could result into negative Lyapunov exponent values, representing stable fixed points. Apparently, the unstable points emergence is due to the extremely low value of $k$.

For the other two fixed points, taking into account figure 2 these appear at the positions: ($\varphi^* = \pm \sqrt{\frac{2|r_0|}{u_0}}, \psi^* = 0$). This furthermore means that the Lyapynov exponents' values are:

$$\lambda_{L_{1,2}} = 1 \pm k(-5|r_0|)^{1/2}, \tag{A7}$$

which are complex numbers with positive real parts, meaning that these points are unstable fixed points.

## B. In Euclidean time

Following the same procedure, in the case of the maps in Euclidean time, described by equations (19) and (20), we calculate Liapunov exponents:

$$\lambda_{L_{1,2}} = 1 \pm k(-|r_0| + 3u_0\varphi^{*2})^{1/2}. \tag{B1}$$

At the critical point $(\varphi^* = 0, \psi^* = 0)$, equation (B1) provides the values $\lambda_{L_{1,2}} = 1 \pm k(-|r_0|)^{1/2}$, which are complex numbers with positive real parts, meaning that the central fixed point is unstable.

Using equation (21) we obtain the following:

$$\lambda_{L_{1,2}} = 1 \pm k(-|r_0| + 3f^2\frac{u_0^2}{r_0})^{1/2}. \tag{B2}$$

By considering $\Lambda > fu_0$, the Lyapunov exponents are becoming complex numbers with positive real parts, meaning that these points are unstable fixed points. But even if this condition is not valid, even very small values of $k$ may render the fixed points unstable ones.